\def\mK{{\rm \mu K}}

\def\expec#1{\langle#1\rangle}

\def\etal{{\frenchspacing\it et al.}}

\def\eg{{\frenchspacing\it e.g.}}
\def\etc{{\frenchspacing\it etc.}}
\def\rms{{\frenchspacing r.m.s.}}

\def\pp{\noindent\parshape 2 0truecm 13.6truecm 1truecm 12.6truecm}
\def\rf#1;#2;#3;#4 {\par\pp#1, {\it #2}, {\bf #3}, #4. \par}
\def\rg#1;#2;#3;#4;#5 {\par\pp#1, {\it #2}, {\bf #3}, #4 (``#5"). \par}
\def\rn{\pp}

\def\beq#1{\begin{equation}\label{#1}}
\def\eeq{\end{equation}}
\def\beqa#1{\begin{eqnarray}\label{#1}}
\def\eeqa{\end{eqnarray}}

\def\bfig{\begin{figure}[h] \centerline{\hbox{}}\vfill}
\def\efig{\end{figure}\vfill\newpage}

\def\fheight{12cm}
\def\fwidth{17cm}
\def\fig#1{Figure~\ref{#1}}

\def\spose#1{\hbox to 0pt{#1\hss}}
\def\simlt{\mathrel{\spose{\lower 3pt\hbox{$\mathchar"218$}}
     \raise 2.0pt\hbox{$\mathchar"13C$}}}
\def\simgt{\mathrel{\spose{\lower 3pt\hbox{$\mathchar"218$}}
     \raise 2.0pt\hbox{$\mathchar"13E$}}}
\def\simpropto{\mathrel{\spose{\lower 3pt\hbox{$\mathchar"218$}}
     \raise 2.0pt\hbox{$\propto$}}}

\def\addr#1{{\small\it #1}}
\def\auth#1{{#1}}

\def\Qrmsps{Q_{rms,ps}}

\def\l{\ell}

\def\expec#1{\langle#1\rangle}

\def\Dl{\Delta\l}

\def\ed{\end{document}}

\documentstyle[11pt,epsf,rotate]{article}
\begin{document}


\begin{titlepage}   

\noindent
\begin{center}

\vskip0.9truecm
{\bf

THE ANGULAR POWER SPECTRUM OF THE 4 YEAR COBE DATA\footnote{
Published in {\it ApJ Lett.} {\bf 464}, L35-L38, June 10, 1996. Submitted Jan. 15, 1996.\\
Available from 
{\it h t t p://www.sns.ias.edu/$\tilde{~}$max/cobepow.html} (faster from the US)\\
and from 
{\it h t t p://www.mpa-garching.mpg.de/$\tilde{~}$max/cobepow.html} (faster from Europe).\\
Note that figures 1, 3 and 4 will print in color if your printer supports it.}
}

\vskip 0.5truecm
  \auth{Max Tegmark}
  \smallskip

  \addr{Max-Planck-Institut f\"ur Physik, F\"ohringer Ring 6,}
  \addr{D-80805 M\"unchen;}

 \addr{email: max@mppmu.mpg.de}

  \smallskip
  \vskip 0.2truecm

\end{center}

\abstract{
The angular power spectrum $C_\l$ is extracted from the
4 year COBE DMR data with a $20^\circ$ galactic cut,
using the narrowest window functions possible. 
The average power in eight multipole bands is also
computed, and plotted together with a compilation
of power spectrum measurements from other experiments.
The COBE results are found to be consistent with the $n=1$
power spectrum with the normalization 
$\Qrmsps=18\,\mK$ reported by the COBE DMR team.
Certain non-standard cosmologies, such as ``small
universe" models with nontrivial spatial topology,
predict power spectra which are not smooth functions.
Rather, they
contain bumps and wiggles that may not have been 
detected by other data analysis techniques such as
the Hauser-Peebles method (Wright {\etal} 1996), the
band-power method (Hinshaw {\etal} 1996) and 
orthogonalized spherical harmonics
(G\'orski {\etal} 1996), since these methods 
all give broader window
functions that can smear such features out beyond recognition.
On the large angular scales probed by COBE, 
the Universe thus appears to be kind to us, presenting a
power spectrum that is a simple smooth function.
}

\end{titlepage}


\section{INTRODUCTION}

Since their discovery (Smoot {\etal} 1992), fluctuations
in the Cosmic Microwave 
Background Radiation (CMB) have emerged as 
possibly one of the most promising
ways of measuring key cosmological parameters such as 
the Hubble constant $h$, 
the density parameter $\Omega$, 
the baryon fraction $\Omega_b$, the cosmological constant $\Lambda$, 
{\etc} 
As the angular power spectrum of the CMB, usually denoted
$C_\l$, depends on about a dozen cosmological parameters
(see {\eg} White {\etal} 1994, Bond 1995a, Hu 1995,
Steinhardt 1995 or Tegmark 1996b 
for recent reviews), accurate 
determination of this power spectrum by future experiments
could be used to measure 
all these parameters simultaneously, 
with accuracies of a few percent or better
(Jungman {\etal} 1996). 
Although we are still years away from 
attaining this goal, which would require mapping the CMB 
at high resolution over a large fraction of the sky, 
our knowledge of the power spectrum has been growing steadily
over the last few years. The first year
of COBE/DMR data (Smoot {\etal} 1992) indicated that the power spectrum 
was approximately scale-invariant $(n\sim 1)$ on the large angular scales
($\l\simlt 30$) to which COBE is sensitive, 
and this conclusion has been confirmed
by the FIRS and TENERIFE experiments 
(Ganga {\etal} 1994; Hancock {\etal} 1994).  
Ground- and balloon-based 
experiments have produced numerous measurements of the fluctuations 
on degree scales $(\l\sim 50-300)$, and although the situation 
is still far from clear, there is now some evidence that the power
spectrum is larger in this $\l$-range than on COBE scales
({\eg} Scott {\etal} 1995; Kogut \& Hinshaw 1996).
Recent results from the CAT experiment 
(Hancock {\etal} 1996; Scott {\etal} 1996)
indicate that the power spectrum has fallen to lower values
at $\l\sim 600$, which could be interpreted as there being a CDM-type
``Doppler peak" around $\l\sim 300$. 

Our knowledge of the power spectrum for $\l\simgt 30$ is 
still quite limited, since the small patches of sky surveyed at high resolution
so far give a large sample variance. On the largest angular scales, however, 
the situation is much better, since COBE has mapped the entire sky.
Indeed, the signal-to-noise ratio in the 4 year COBE data 
(Bennett {\etal} 1996)
is so good that 
the error bars on the 
$C_\l$-estimates for $\l\simlt 10$ are now entirely dominated by cosmic variance.
This means that (apart from reducing possible systematic errors and 
correcting for residual foreground contamination), 
these are the best estimates that
mankind will ever be able to make of the large-scale power. 

In view of this experimental progress, it is clearly
worthwhile to estimate the power spectrum from the 4 year COBE 
data as accurately as possible. This is the purpose of the present 
{\it Letter}.

\subsection{Power spectrum estimation with incomplete sky coverage}

Pioneering work on the problem of power spectrum estimation
from experimental data
(Peebles 1973; 
Hauser \& Peebles 1973) has recently been extended and applied
to the 4 year COBE data (Wright {\etal} 1996, hereafter W96). 
When estimating power spectra,
it is customary to place both vertical and horizontal error bars
on the data points, as in {\eg} \fig{ExperimentsFig}. 
The former represent
the uncertainty due to noise and sample variance, and the latter 
reflect the fact that an estimate of $C_\l$ inadvertently also 
receives contributions from other multipole moments. 
In other words, the estimate of $C_\l$ is in fact a weighted average
of a band of multipoles. The weights are referred to as the
{\it window function}, and for the estimate of $C_\l$ to be
a good one, we clearly want the window function to be
centered on $\l$ with an {\rms} width $\Delta\l$ that is as small
as possible. 
As is well-known, incomplete
sky coverage destroys the orthogonality of the
spherical harmonics, and makes it impossible to 
attain perfect spectral resolution, $\Dl=0$.
For typical ground- and balloon-based experiments probing degree
scales, the relative spectral blurring $\Dl/\l$ tends to be of order unity, 
which makes it difficult to resolve details such as the number of Doppler
peaks. A much better method is that used in W96, where 
the relative spectral resolution $\Dl/\l$ is brought down to the order 
of $25\%$ by using spherical harmonics. In Tegmark (1996a,
hereafter T96), a method was
presented for 
reducing these horizontal 
error bars still further, down to their theoretical 
minimum, which for a $20^\circ$ galactic cut was seen to 
be $\Dl\sim 1$. 

\subsection{The importance of high spectral resolution}

Just as high spectral resolution is crucial in for instance absorption
line studies (since it prevents interesting features from getting 
smeared out), 
it is also important when measuring the CMB power spectrum. 
The reason is that we cannot a priori assume that the power spectrum will
be a simple smooth function. Indeed, there are 
non-standard cosmologies, such as ``small
universe" models with nontrivial spatial topology,
that predict power spectra which on average have an
$n\sim 1$ slope but contain bumps and wiggles that can 
only be resolved with a high spectral resolution 
(Stevens {\etal} 1993; de Oliveira-Costa \& Smoot 1995).
In addition, lowering the spectral resolution degrades information:
broad window functions make the estimates of 
nearby multipoles highly correlated, so that the resulting
power spectrum plot will contain fewer independent data
points than one may naively expect.

\bigskip

The remainder of this {\it Letter} is organized as follows. 
In Section 2, the four year COBE data set is analyzed. 
In Section 3, the
results are discussed and compared with those obtained with other methods
such as the 
orthogonalized
spherical harmonic method of G\'orski (1994, hereafter G94)
and the
signal-to-noise eigenmode methods of 
Bond (1995b, hereafter B95) 
and Bunn and Sugiyama (1995, hereafter BS95). 

\section{RESULTS}

\subsection{The method}

How to extract the angular power spectrum from a CMB
map with maximum spectral resolution is been described in detail in T96.
We will apply this method to the 4 year data here exactly as it was applied
to the 2 year COBE data in T96. 
Below we give merely a brief review of how the method works, referring 
the interested reader to T96 for technical details.

A simple estimate of 
the multipole $C_\l$ is obtained by taking a linear combination of
all pixels, squaring it, and subtracting off the 
expected noise contribution. The weights in the linear combination
are conveniently plotted as a sky map in the same way that we plot the data, 
and we usually refer to the set of weights as a {\it weight function}. 
It is easy to show that the expectation value of such an estimate
is a linear combination of all the true multipoles. The coefficients in this
linear combination are called the {\it window function}, a function of $\l$
(see the examples in \fig{WindowFig}).
In other words, given any weight function, there is a corresponding window function.
The window function turns out to be simply the square of the
spherical harmonic coefficients of the weight function, summed over $m$.

In the Hauser-Peebles method, the weight functions are chosen to be the
spherical harmonics, but set equal to zero inside the galaxy cut and appropriately rescaled.
A number of other weight functions have been employed in CMB analyses,
for instance the orthogonalized spherical harmonics of 
G94 and the signal-to-noise
eigenfunctions of B95 and BS95
(these functions were tailor made for the problem 
of efficient parameter estimation with likelihood analysis, not for
power spectrum estimation).
The method of T96 simply employs those weight functions that give the 
{\it narrowest window functions possible}, and it is shown that 
these functions can be found by
solving a certain eigenvalue problem numerically. 
In the simplest version of 
the method (the version we use here --- it has $\gamma=0$ as in T96), 
these functions are
completely independent of any prior assumptions about the power spectrum, and
are hence determined by the geometry of the galaxy cut alone.

What are these weight functions like?
A sample weight function is plotted in T96 together with the corresponding 
spherical harmonic, and it is seen that they look quite similar far away from
the galactic cut. The main difference is that the optimal weight functions
approach zero smoothly at the edge, whereas a truncated spherical harmonic does
not. As is discussed in further detail in T96, this absence of sharp edges in 
the weight functions is a key feature of the method, since sharp edges cause
``ringing" in Fourier space (in the multipole domain), which corresponds to an
unnecessarily wide window function.

To reduce error bars, a multipole is estimated as a weighted average of estimates
of the above-mentioned simple type, just as in the Hauser-Peebles method where 
$C_\l$ is estimated by an average of the square amplitudes of the
$(2\l+1)$ different spherical harmonic coefficients.


\subsection{The data}

The 53 and 90 GHz channels (A and B) of the
COBE DMR 4 year data were combined into a single sky map 
by the standard minimum-variance weighting, pixel by pixel. 
We use the data set that was pixelized in galactic coordinates.
After removing all pixels less than $20^\circ$ away from the 
galactic plane, 4016 pixels remain.
As has become standard, we make no attempts  
to subtract galactic contamination outside the cut.

The resulting power spectrum is shown in \fig{ClFig}.
A brute force likelihood analysis of the 4 year data
set (Hinshaw {\etal} 1996) gives a best fit normalization 
of $\Qrmsps=18.4\,\mK$ for a simple $n=1$ model,\footnote{
By this we mean an $n=1$ model including only the
Sachs-Wolfe effect, so that $C_\l\propto 1/\l(\l+1)$. 
Note that the slow rise towards the first Doppler peak
in an $n=1$ CDM model gives a best fit 
Sachs-Wolfe spectrum with $n\sim 1.15$, whereas 
$n=1$ models with spatial curvature or cosmological 
constant can give best fit 
Sachs-Wolfe spectra with $n<1$.
}
corresponding to the
heavy horizontal line in the figure.
If this model were correct, we would expect approximately 
$68\%$ of the data points to fall within the shaded
$1-\sigma$ error region.
As can be seen, the height of this region 
(the size of the vertical error bars) 
is dominated by cosmic variance 
for low $\l$ and by noise for large $\l$.
At the cost of increasing $\Dl$, the variance 
can of course be reduced further by 
grouping multipoles together in bands and averaging them with 
minimum-variance weighting,
as shown in figures~\ref{ExperimentsFig} 
and~\ref{DopplerFig}
and in Table 1. 
We have followed W96 and chosen the eight 
multipole bands 2, 3, 4, 5-6, 7-9, 10-13, 14-19 and 20-30.

\begin{table}
$$
\begin{tabular}{c|ccccc}
\hline
Band&$\expec{\l}$&$\Delta\l$&$\delta T$&$-1\sigma$&$+1\sigma$\\
\hline
1&2.31&0.89& 9.6& 0.0&22.6\\
2&3.45&1.06&25.4&15.7&32.3\\
3&4.38&1.26&27.5&20.7&32.9\\
4&5.87&1.43&28.1&24.0&31.7\\
5&8.34&1.62&25.9&22.1&29.3\\
6&11.9&1.82&22.3&17.8&26.1\\
7&17.0&2.26&30.4&25.3&34.7\\
8&25.6&3.49&31.8& 0.0&48.6\\
\hline
\end{tabular}
$$
\caption{The COBE power spectrum 
$\delta T\equiv [\l(\l+1)C_\l/2\pi]^{1/2}$ in $\mu K$.}
\label{Table1}
\end{table}

For verification, 1000 mock COBE maps were generated for the 
model $n=1$, $\Qrmsps=18.4\,\mK$ and piped through the data analysis software.
As expected, the extracted multipoles were found to be unbiased estimates of the 
true multipoles, and the scatter was in agreement with
the (analytic) error bars shown in
the figures.

\subsection{The window functions}

The horizontal bars in \fig{ClFig} are seen to be fairly independent of $\l$, just
as expected --- as discussed in detail in T96, 
the angular size of the two sky
patches surviving the galaxy cut is $\Delta\theta\sim 1$ radian, 
and we expect $\Delta\l \sim 1/\Delta\theta\sim 1$. 
A typical window function is shown in 
\fig{WindowFig}, and exhibits the following features that are common to
all our window functions:
\begin{itemize}
\item For even $\l$, the window function vanishes for all odd 
multipoles, and vice versa. This happens because the galactic cut 
is symmetric about the galactic plane, and thus preserves the orthogonality
between even and odd spherical harmonics, since they have opposite parity.
\item The window function is for all practical purposes zero for 
multipoles below $\l-2$ and above $\l+2$. 
\item The central value is typically about three times as high as 
the two sidelobes.
\end{itemize}
The multipole $D_\l\equiv\l(\l+1)C_\l$ is thus typically estimated by something like
$0.6 D_\l + 0.2 D_{\l-2} + 0.2 D_{\l+2}$.
This corresponds to all data points in \fig{ClFig}
being uncorrelated, with the exception that 
points separated by $\Dl=2$ have a correlation of order $55\%$ and 
points separated by $\Dl=4$ have a correlation of order $9\%$.
Note that neighboring points are completely uncorrelated. 

The only exception to the above is the window function for the quadrupole,
$\l=2$: since it is required to vanish at $\l=0$, it picks up a non-negligible
contribution from $\l=6$ instead.

\section{DISCUSSION}

We have computed the angular power spectrum of the 4 year COBE DMR data
using the maximum resolution method of T96.
The signal-to-noise ratio in the data is now so high that 
the error bars for $\l\simlt 10$ are entirely dominated by cosmic
variance. This means that, apart from future corrections due to better modeling
of foregrounds and systematics, this is close to the best measurement of the low
multipoles that mankind will ever be able to make, since cosmic variance could only be reduced 
by measurements in a different horizon volume. 

\begin{figure}[phbt]
\centerline{{\vbox{\epsfxsize=14cm\epsfysize=14cm\epsfbox{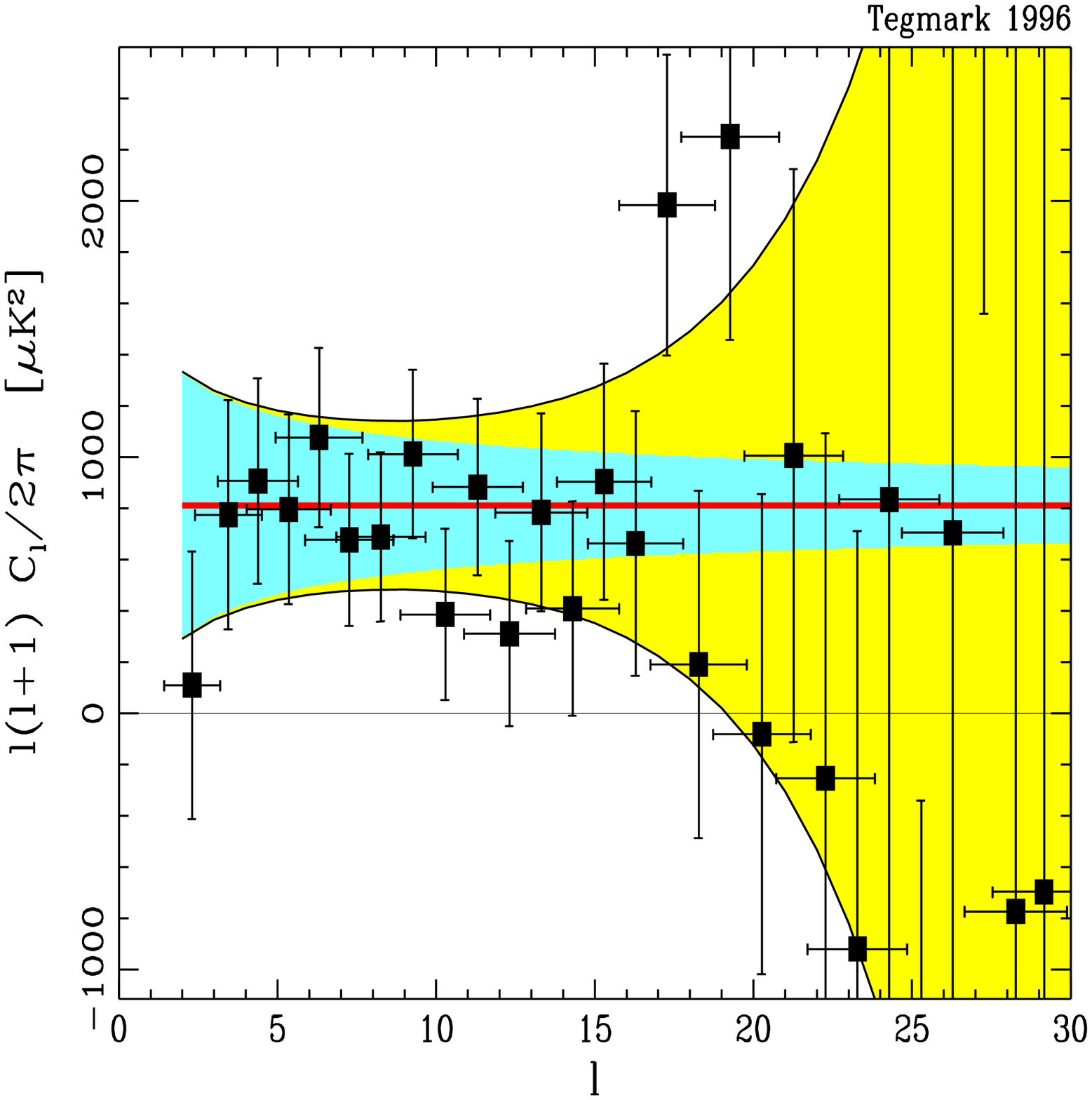}}}}
\vskip-2cm
\caption{
The power spectrum observed by COBE.
}
The observed multipoles $D_\l = \l(\l+1)C_\l$ are plotted with
$1-\sigma$ error bars. The vertical error bars include both
pixel noise and cosmic variance, and the horizontal bars 
show the width of the window functions used. 
If the true power spectrum is given by 
$n=1$ and $Q_{rms,ps}=18.4\mK$ (the heavy horizontal line), 
then the shaded region gives the $1-\sigma$ error bars and
the dark-shaded region shows the contribution from
cosmic variance. 
\label{ClFig}
\end{figure}

The power spectrum in \fig{ClFig}
is seen to be consistent with an 
$n=1$, $\Qrmsps=18.4\,\mK$ model.
This model is close to the best-fit models found in
the various two-parameter Bayesian likelihood analyses
(\eg, Hinshaw {\etal} 1996, G\'orski {\etal} 1996, W96), which we can interpret
as the best-fit straight line through the data points
in \fig{ClFig} being close to the
horizontal heavy line.
As has frequently been pointed out (see {\eg} White \& Bunn 1995),
Bayesean methods by their very nature can only make
{\it relative} statements of merit about different models,
and never address the question of whether the best-fit model
itself is in fact inconsistent with the data.
As an absurd example, the best fit straight line to a
parabola on the interval $[-1,1]$ is horizontal, even though
this is a terrible fit to the data. It is thus quite
reassuring that the power spectrum in \fig{ClFig}
not only has the right average normalization and
slope, but that each and every one 
of the multipoles appear to be consistent with this standard 
best fit model.\footnote{Kogut {\etal} (1996)
find that correcting for galactic 
foreground emission increases the value of the CMB quadrupole, 
making it consistent with the best fit power spectrum.}

\subsection{Comparison with other results}

\begin{figure}[phbt]
\centerline{\rotate[r]{\vbox{\epsfysize=14cm\epsfysize=14cm\epsfbox{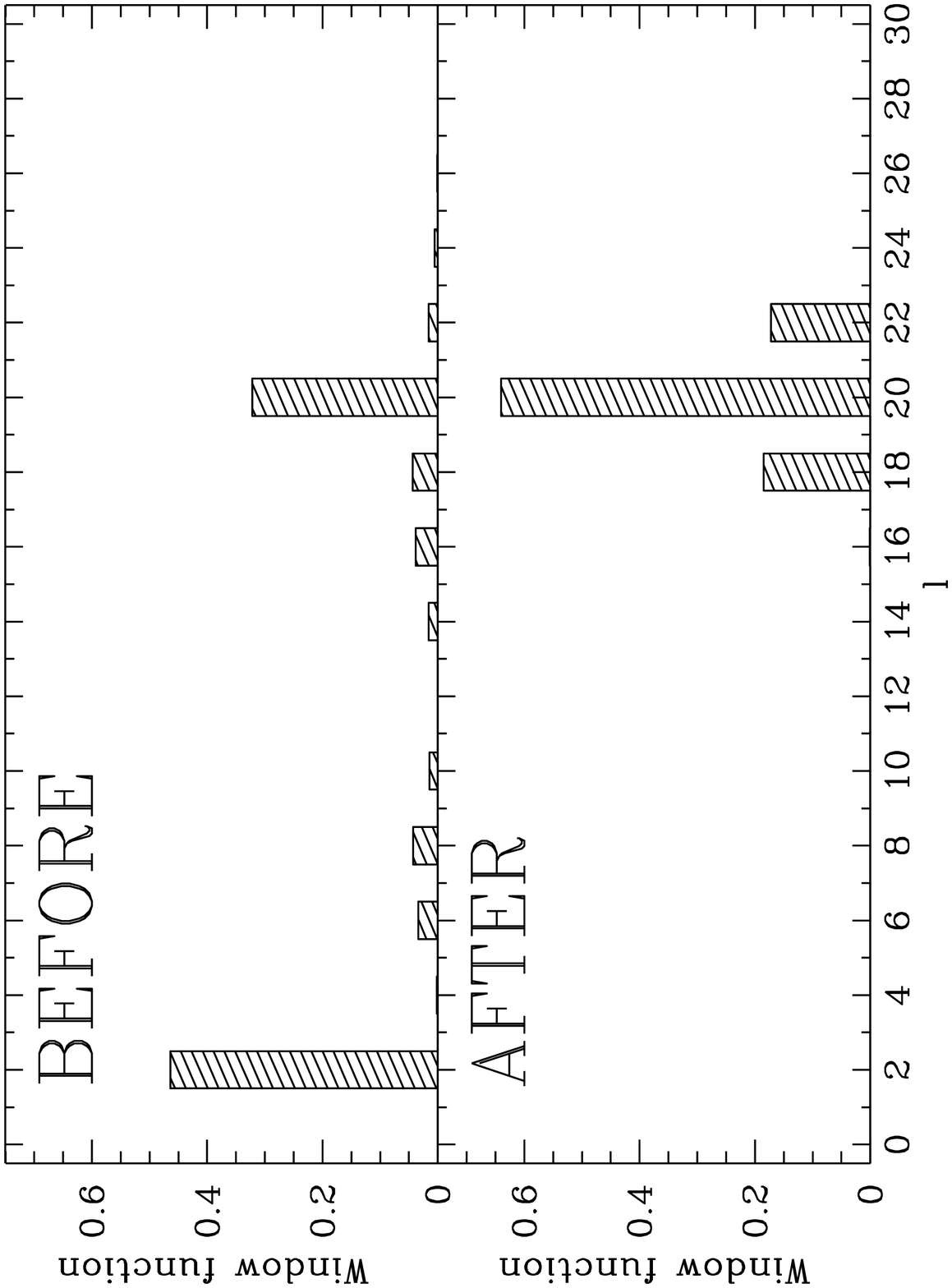}}}}
\caption{
Window functions before and after optimization.
}
Two window functions for estimation of the multipole
$\l=20$, $m=0$ are shown. The upper one is that of the
spherical harmonic method (W96), which exhibits a strong leakage
from lower multipoles such as the quadrupole.
The lower one is the one resulting from the optimal method.
\label{WindowFig}
\end{figure}

\begin{figure}[phbt]
\centerline{{\vbox{\epsfxsize=14cm\epsfysize=14cm\epsfbox{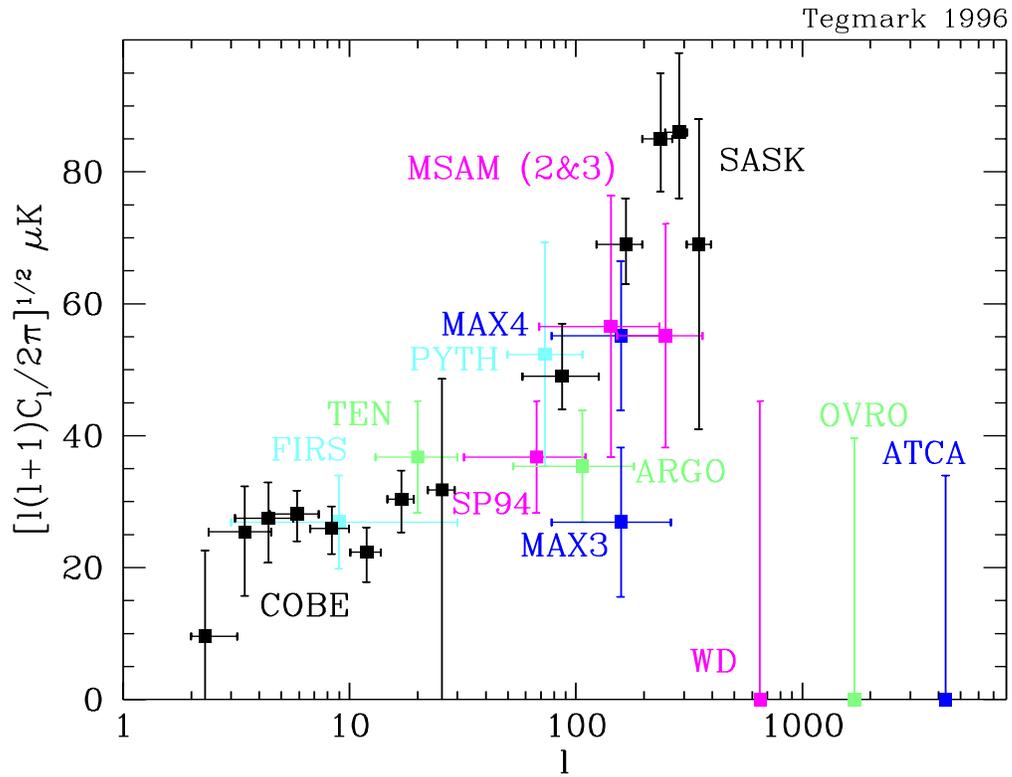}}}}
\vskip-2cm
\caption{
Power spectrum measurements.
}
The observed multipoles $D_\l = \l(\l+1)C_\l$ are plotted 
for a selection of experiments.
Both vertical and horizontal 
bars have the same meaning as in \fig{ClFig}. 
The COBE
data are those of \fig{ClFig}, averaged over 8 multipole bands,
and the rest are from the Saskatoon 
experiment (Netterfield {\etal} 1996)
and from the compilation of Scott {\etal} (1995). 
\label{ExperimentsFig}
\end{figure}

\begin{figure}[phbt]
\centerline{{\vbox{\epsfxsize=14cm\epsfysize=14cm\epsfbox{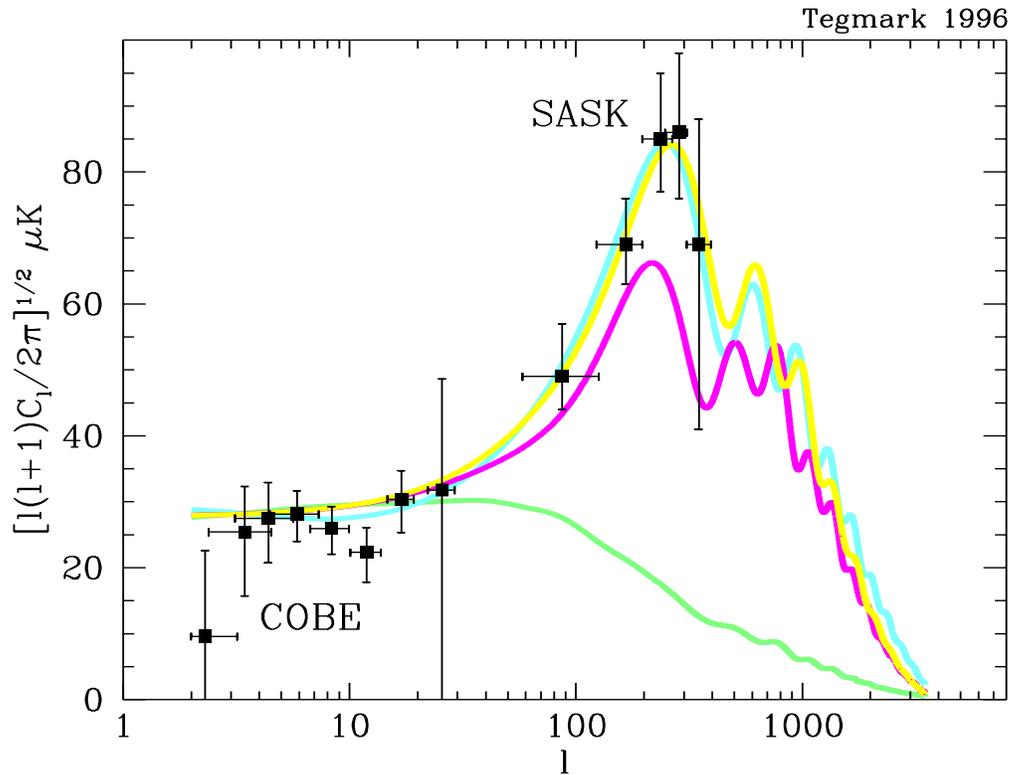}}}}
\vskip-2cm
\caption{
Power spectrum models.
}
The data points from \fig{ExperimentsFig} with the narrowest 
window functions (COBE and Saskatoon) 
are compared with the predictions from
four variants of the standard CDM model from Sugiyama (1995),
all with $n=1$ and $\Omega_b=0.05$.
From top to bottom at $\l=200$, they are
a flat model with $\Lambda=0.7$, 
a model with $h=0.3$, the standard $h=0.5$ model and 
a model with a reionization optical depth $\tau\sim 2$.
\label{DopplerFig}
\end{figure}

A number of other linear techniques for CMB analysis have recently been
applied to the COBE data. 
Both 
the orthogonalized spherical harmonics method
(G94),
the Karhunen-Lo\`eve (KL) signal-to-noise eigenmode method
(B95, BS95), 
and the brute-force method 
(Tegmark \& Bunn 1995, Hinshaw {\etal} 1996)
were devised to solve a different problem than the one
addressed here.
If one is willing to parametrize the power spectrum by a small
number of parameters, for instance a spectral index and an amplitude,
then these methods provide an efficient way of estimating
these parameters via a likelihood analysis.
Why cannot the basis functions of these methods be used to
estimate the angular power spectrum $C_\l$ directly, as
they are after all orthogonal over the galaxy-cut sky?
The answer is that these basis functions are orthogonal to
{\it each other}, whereas in our context, we want them to be
as orthogonal as possible not to each other but to the
{\it spherical harmonics}. This is illustrated in
\fig{WindowFig}, which contrasts $\l=20$ window functions
of the optimal method and the generalized Hauser-Peebles method
(W96, de Oliveira-Costa \& Smoot 1995).
We want the window function to be centered on $\l=20$, and
be as narrow as possible, so the lower one is clearly preferable.
The upper weight function is
seen to couple strongly to many of the lower multipoles,
and picks up a contribution from the quadrupole that is even
greater than that from $\l=20$. This of course renders 
it inappropriate
for estimating the power at $\l=20$.
Analogous window functions can readily be computed for
the orthogonalized spherical harmonics of G94 or
the signal-to-noise eigenmodes of B95 and BS95.
They are also
broader than the optimal one in \fig{WindowFig} --- the
optimal weight functions of course
give narrower window functions than other basis functions 
{\it by definition}, since they were defined as those functions
that give the narrowest window functions possible.

It should be emphasized that generating such window functions
for the basis functions of G94, B95 and BS95 would be quite an
unfair criticism of these methods, since this would be grading them
with respect to a property
that they were not designed to have.
These authors have never claimed that their basis functions were optimal
for multipole estimation, merely (and rightly so) that they were
virtually optimal for parameter fitting with a likelihood analysis.

The power spectrum of \fig{ClFig} is also consistent with that 
extracted in W96 using the Hauser-Peebles method.
At low $\l$, the individual multipoles estimates agree well with each
other. As $\l$ increases, the spectral resolution $\Dl$ 
of the Hauser-Peebles method grows approximately linearly (see T96) whereas
the resolution in \fig{ClFig} is seen to more or less remain constraint.
Thus the data points in $W96$ begin to differ from those in \fig{ClFig}
at larger $\l$, since the former are no longer probing individual
multipoles but weighted averages of a broad range.

\smallskip

In summary, the 4 year COBE data has measured 
the power spectrum for $\l\simlt 15$ with an accuracy
approaching the cosmic
variance limit.
As the next generation of CMB experiments extend this 
success to smaller angular scales, 
the CMB may turn out to be one of the most potent arbiters
between cosmological models.

\bigskip
The author wishes to thank Ang\'elica de Oliveira-Costa, 
Krystof G\'orski, Ned Wright and an anonymous referee for helpful 
comments on the manuscript.
This work was partially supported by European Union contract
CHRX-CT93-0120 and Deutsche Forschungsgemeinschaft grant
SFB-375. The COBE data sets were developed by the NASA
Goddard Space Flight Center under the guidance of the COBE Science Working 
Group and were provided by the NSSDC.


\clearpage

\section{REFERENCES}

\rf Bennett, C. L. {\etal} 1996;ApJ;464;L1

\rn Bond, J. R. 1995a, in 
{\it Cosmology and Large Scale Structure}, ed. Schaeffer, R. (Elsevier).

\rg Bond, J. R. 1995b;Phys. Rev. Lett.;74;4369;B95

\rg Bunn, E. F. \& Sugiyama, N. 1995;ApJ;446;49;BS95

\rf de Oliveira-Costa, A. \& Smoot, G. F. 1995;ApJ;448;477

\rf Ganga, K. {\etal} 1994;ApJ;432;L15

\rg G\'orski 1994;ApJ;430;L85;G94

\rf G\'orski {\etal} 1996;ApJ;464;L11

\rf Hancock, S. {\etal} 1994;Nature;367;333

\rn Hancock, S. {\etal} 1996, submitted to {\it Nature}.

\rf Hauser, M. G. \& Peebles, P. J. E. 1973;ApJ;185;757

\rn Hinshaw, G. {\etal} 1996;464;L17
     
\rn Hu, W. 1995, in {\it The Universe at High-z}, eds. E. Martinez-Gonzalez and 
J.L. Sanz (Springer, in press), astro-ph/9511130.

\rn Jungman {\etal} 1996, preprint astro-ph/9512139.

\rf Kogut, A. {\etal} 1996;ApJ;464;L5
     
\rf Kogut, A. \& Hinshaw, G. 1996;ApJ;464;L39

\rn Netterfield, C. B. {\etal} 1996, preprint astro-ph/9601197.

\rf Peebles, P. J. E. 1973;ApJ;185;413

\rf Scott D, Silk J and White M 1995;Science;268;829

\rn Scott {\etal} 1996, in preparation.

\rf Smoot, G.F. {\etal} 1992;ApJ;396;L1


\rn Steinhardt P S 1995, preprint astro-ph/9502024.

\rf Stevens, D., Scott, D. \& Silk, J. 1993;Phys. Rev. Lett.;71;20

\rf Sugiyama, N. 1995;ApJS;100;281

\rf Tegmark, M. \& Bunn, E. F. 1995;ApJ;455;1

\rf Tegmark, M. 1996a;MNRAS;280;299 (``T96").

\rn Tegmark, M. 1996b, to appear in {\it Proc. Enrico Fermi, Course CXXXII, Varenna},
eds. Bonometto, S. \& Primack, J., astro-ph/9511148.

\rf White, M. \& Bunn, E. F. 1995;ApJ;450;477

\rf White, M., Scott, D. and Silk, J. 1994;ARA\&A;32;319

\rg Wright, E. L. {\etal} 1996;ApJ;464;L21;W96
 
\def\fheight{10.3cm} \def\fwidth{14.5cm}
 
\clearpage

\clearpage

\clearpage

\clearpage

\end{document}